\documentclass[10pt,journal,compsoc]{IEEEtran}

\usepackage[keys, probability, sets, logic, operators, adversary, 
   ff, primitives, advantage]{cryptocode}

\usepackage{graphicx,amsthm} 
\createprocedurecommand{gamer}{}{}{linenumbering}
\NewEnviron{game}[1][]{\noindent\gamer{#1}{\BODY}}

\newcommand{\atime}{\mathtt{time}}
\newcommand{\atype}{\mathtt{type}}
\newcommand{\adata}{\mathtt{data}}
\newcommand{\atags}{[\mathtt{tags}]}
\newcommand{\counting}{\mathtt{count}}

\newcommand{\ltime}{\mathtt{time}}

\newcommand{\signature}{\sigma}

\newcommand{\cdata}{\mathtt{cdata}}

\usepackage{pgf-umlsd}
\usepackage{tikz}
\usepackage{adjustbox}
\usepackage{xspace}

\ifCLASSOPTIONcompsoc
  \usepackage[nocompress]{cite}
\else
  \usepackage{cite}
\fi
\ifCLASSINFOpdf
\else
\fi
\begin{document}

\title{TAPESTRY: A Blockchain based Service for Trusted Interaction Online}

\author{Yifan Yang, Daniel Cooper, John Collomosse, Constantin  C. Dr\u{a}gan, Mark Manulis, \\Jamie Steane, Arthi Manohar, Jo Briggs, Helen Jones, Wendy Moncur
        
\IEEEcompsocitemizethanks{\IEEEcompsocthanksitem Y. Yang, D. Cooper and J. Collomosse are with the Centre for Vision Speech and Signal Processing (CVSSP), University of Surrey, UK.
\IEEEcompsocthanksitem C. C. Dr\u{a}gan and M. Manulis are with the Surrey Centre for Cyber-Security (SCCS), University of Surrey, UK.
\IEEEcompsocthanksitem J. Steane and J. Briggs are with the School of Art and Design, Northumbria University, UK.  A. Manohar was with Northumbria University at the time this work was undertaken and is now at Brunel University, UK.
\IEEEcompsocthanksitem W. Moncur is  with the Duncan Jordanstone Centre for Art and Design (DJCAD), University of Dundee, UK. H. Jones was with DJCAD at the time this work was undertaken and is now at UCLan, UK.\protect\\

Corresponding Author: yifan.yang@surrey.ac.uk.}
\thanks{Manuscript received May 9, 2019.}}

\markboth{Preprint Submitted to IEEE Transactions on Services Computing, 9~May~2019}%
{Yang \MakeLowercase{\textit{et al.}}: A Blockchain based Service for Trusted Interaction Online}

\IEEEtitleabstractindextext{%
\begin{abstract}
We present a novel blockchain based service for proving the provenance of online digital identity, exposed as an assistive tool to help non-expert users make better decisions about whom to trust online.  Our service harnesses the digital personhood (DP); the longitudinal and multi-modal signals created through users' lifelong digital interactions, as a basis for evidencing the provenance of identity. We describe how users may exchange trust evidence derived from their DP, in a granular and privacy-preserving manner, with other users in order to demonstrate coherence and longevity in their behaviour online.  This is enabled through a novel secure infrastructure combining hybrid on- and off-chain storage combined with deep learning for DP analytics and visualization. We show how our tools enable users to make more effective decisions on whether to trust unknown third parties online, and also to spot behavioural deviations in their own social media footprints indicative of account hijacking.  
\end{abstract}

}

\newcommand{\etal}{et~al.\xspace}
\newcommand{\eg}{e.\,g.\xspace}
\newcommand{\ie}{i.\,e.\xspace}
\newcommand{\squeezeupSmall}{\vspace{-2mm}}
\newcommand{\squeezeup}{\vspace{-4mm}}

\renewcommand{\baselinestretch}{0.95}

\maketitle

\IEEEdisplaynontitleabstractindextext

%
\IEEEpeerreviewmaketitle



%
%
%
%

\IEEEraisesectionheading{\section{Introduction}\label{sec:introduction}}

\IEEEPARstart{O}{nline} fraud and scams are sharply on the increase, costing the global economy in excess of US\$3 trillion in 2018 \cite{CostFraud2018} and are often perpetrated through ephemeral false identities.  Users struggle to make decisions on who to trust online, exposing themselves to risks from inappropriate over-disclosure of personal data.  This motivates new techniques for determining the provenance and trustworthiness of the digital identities --- people, business or services ---  encountered online.

This paper reports a novel blockchain-based service that harnesses the complex longitudinal and multi-modal signals within citizens' digitally mediated interactions (for example, on social media) to support safe online interactions.  These signals created - photos shared, comments left, posts `liked' etc. - weave a complex `tapestry' reflecting our relationships, personality and identity, referred to as the `Digital Personhood' (DP). Commodification of the DP now fuels a billion-dollar industry in which machine learning is increasingly utilised to help make sense of, and extract value from, the deluge of DP data.  In this work we exploit the DP for the social good; through a platform (herein referred to as 'TAPESTRY') that empowers users to share `trust evidence' of their DP in a granular, privacy preserving manner, in order to prove the provenance of their digital identity and so engender trust online.   

We draw distinction between the problem of proving identity (authentication), and the problem tackled here, of proving the provenance of a digital identity.  Online security is typically reliant on traditional representations of identity, taking simple pseudonyms or email addresses “at face value” to interact with one another or digital services. We are now entering a new era in which citizens will construct a DP from childhood, comprising rich lifelong digital trails from social media and interactions with technology~\cite{Orzech-etal2018}. Those accumulated signals offer an increasingly viable way to prove the veracity of a digital identity. Leveraging the DP for this purpose poses significant challenges around signal processing, privacy, information security and infrastructure. Further challenges arise by designing the service for non-experts, who may have low levels of digital literacy - especially around numeracy.  A fundamental tenet of the TAPESTRY platform is the preservation of the end-user as the owner of trust decisions; we do not wish to develop a `trust traffic light' or trust scoring system.  Rather we wish to summarise in an intuitive way the trust evidence disclosed from one user to another, in order to support strong decision making around trust using that evidence. TAPESTRY tackles these challenges through three novel technical contributions:
\begin{enumerate}
\item{A {\bf secure data architecture} combining off-chain storage of encrypted trust evidence derived from the DP, backed by an unpermissioned proof-of-work (PoW) blockchain to ensure the integrity and provenance of that evidence.  The architecture incorporates a symmetric key sharing scheme, enabling granular disclosure of trust evidence.  This provides users with agency over to whom evidence is disclosed, as well as control over the time periods and kinds of DP activity disclosed. (Secs.~\ref{sec:platform},~\ref{sec:crypto}).}
\item{A {\bf machine learning} (ML) algorithm to irreversibly gist DP activity into
compact descriptors that serve as the basic unit of trust evidence for sharing in the platform.  We propose a deep neural network (DNN) to extract this evidence through a combination of semantic embedding and temporal modelling, enabling behavioural deviation to be detected over time.  This in turn enables quantification of the regularity and temporal coherence of trust evidence which, combined with assurances over provenance and integrity from the blockchain, serves as the basis for users to make better trust decisions (Sec.~\ref{sec:ml}).  }
\item{A {\bf data visualisation} technique for representing the regularity and coherence of the trust evidence disclosed by a user within a single static image.  The design of the visualisation is evaluated and shown to enable non-expert users to quickly make accurate determinations of the trustworthiness of a digital identity previously unknown to them (Sec.~\ref{sec:viz}).}
\end{enumerate}

In order to evaluate our technical prototype of the  TAPESTRY service, we explore two user-centric case studies where valid trust judgements and the avoidance of either fraud or victimisation are desirable for users.

First, we explore the efficacy of TAPESTRY to help users to detect fraudulent profiles in the context of crowdfunding within the video games industry.  Crowdfunding is a common vehicle by which small studios obtain financial support for new projects, and an online interaction in which investors consider the trustworthiness of pitchers as a primary factor in investment.
We developed a controlled, workshop based evaluation of TAPESTRY in which the platform was used as an aid to investment decision making within a mocked-up crowdfunding scenario. In this scenario, we used TAPESTRY to visualize trust evidence derived from real-world social media footprints of games developers, and artificial profiles fabricated and curated in months prior to the study. We show that whilst TAPESTRY users do not make materially different trust decisions in terms of their accuracy (distinguishing the provenance of fake versus real identities), they are able to do so significantly more quickly using TAPESTRY leading to advantages when making decisions online in time pressured information overloaded situations.

Second, we explore the efficiency of TAPESTRY to help users detect unusual patterns of behaviour within their own social media profile pointing to unauthorized use (or account 'hijacking').  Again our goal is not to automatically raise an alarm or classify this behaviour, but to visually gist the trust evidence derived from a social media profile.  When accustomed to the visual 'look and feel' of TAPESTRY visualisations of this trust evidence, we show that users can perceive deviations from the norm and so spot unusual patterns in online activities posted under their profile.

\section{Related Work}\label{sec:relatedwork}

Open authentication models (e.g. OAuth2) exist for establishing cross-site login without credential sharing, relying upon a trusted identity provider (e.g. Google, Facebook) to approve access to a digital identity. In addition, there is a range of services which help to establish trust, for both named and anonymous/pseudonymous users. For example, Escrow is a contractual arrangement used within the Dark Web, facilitated via a third party, which engenders trust between buyer and seller for crypto-currency transactions~\cite{Tzanetakis-etal2016}. However, TAPESTRY is not proposing yet another access control solution or service for hosting digital identities or the DP, or for facilitating trusted transactions. Rather, TAPESTRY proposes an entirely new kind of service through which one may verify the trustworthiness of a digital identity through evidence derived from signals within an identity's DP.  

\subsection{Signals for Online Trust}
The nature of trust is complex. It is '...a psychological state comprising the intention to accept vulnerability based upon positive expectations of the intentions or behaviours of another'~\cite{Rousseau1998}. Importantly, this definition positions risk as naturally co-existing with trust: an individual accepts that the other party in an interaction may or may not act in the expected manner, but believes that their intentions are good. Offline, judgments about trust are informed by routinely available emotional and behavioural cues \cite{Rocco1998,Cheshire2011,Hancock2015}. Online, these cues are usually absent. However, recent work indicates that there are alternate factors that may inform trust judgments \cite{Jones2018}.

In the context of our crowdfunding case study of TAPESTRY (c.f. Sec.7.2), prior work indicates that these factors include (i) ‘herding’, where (e.g.) potential investors are reassured by the behaviour of previous investors, on the assumption that if others are doing something, it must be the rational thing to do~\cite{Cialdini2001,Kuppuswamy2013} and (ii) ‘social proof’, where (e.g.) less-expert investors are encouraged to invest later in a campaign by the involvement of early investors who are experts in product development or financial investment~\cite{kim-viswanathan2014}. A further factor is social engagement. For example, trust is generated when creators of a crowdfunding campaign provide investors with updates on positive progress towards published goals. This reassures investors and – indicative of trust - increases their investment~\cite {Hornuf2015,Kuppuswamy2013}. Trust is also generated when creators link their social media accounts~\cite{Vismara2016}: investors likely feel that the creator has nothing to hide. Although such observations exist in the literature, TAPESTRY is unique in aggregating evidence from such sources to aid the user in their decision making on trust.

\subsection{Social Media Analytics for Trust}

Use of online services now pervades society and automated profiling of the digital footprint is  increasingly used for identity verification online~\cite{yampolskiy2008behavioural}.  Digital Personhood-related research increasingly involves exploring the behaviours and activities of an online user through the large volumes of data a user generates through \eg social media posts and search histories that collectively comprise their digital footprint~\cite{kosinski2013private}. Such footprints play a crucial role across many digital economy services including user profiling~\cite{farnadi2018user},  personality~\cite{lambiotte2014tracking} and crowdfunding~\cite{nevin2017social}. Therefore, it's vitally important to protect DP by early detecting any malicious activities in social media feeds to prevent economic or reputational loss. There exists various research methods in social media analytics for trust. Chalapathy \etal \cite{chalapathy2018group} and Yu \etal \cite{yu2015glad} explore abnormal behaviours detection from regular group patterns, while Kang \etal \cite{kang2014big} use relationship graph to model social network activities to detect anomalous events. In~\cite{akoglu2013opinion}, the authors use social graph and text information to detect fraudulent comments in online review systems. Phua \etal \cite{phua2010comprehensive} focus on structural metadata within the social activities instead of content.  However, modeling users' behavioural norms in social media over longitudinal time periods, as well as visually representing this analysis to end-users, remains an open challenge that our research aims to address. 

\subsection{Trust and Identity over Blockchain}
Blockchain's innovation is in its facilitation of direct transfer of unique digital property (e.g. currency, data, certificates) - previously reliant on third party intermediaries~\cite{Elsden2018,Andreesen2014}. This promotes ‘trustless trust’~\cite{Andreesen2014} whereby exchanges are `unidirectionally' trustworthy and interpersonal trust, or trust in another online agent is replaced through the technology's functionality of transparency, codification and immutability as a cryptographic audit trail~\cite{Elsden2018}. However, while Blockchain brings significant trust-related functionality for end users many questions remain, both about how to demonstrate and prove such trust to an end user, and if - and if so how - a blockchain service enables this over existing intermediaries.  Elsden et al. ~\cite{Elsden2018} catalogued over 200 blockchain applications and found that amongst identity management systems, most digital identities were provided by a third party (e.g Facebook or email account) or required supplementary state-backed documentation (passport, social security numbers etc.) to prove an identity. Amongst self-sovereign digital identity Blockchain services, where an individual issues and controls their own identity, many involved biometrics (e.g. fingerprint or iris scan) supported by other personal information (email address, bank details etc.). Dunphy and Petitcolas's in-depth review of identity management models using DLT~\cite{Dunphy&Petitcolas} also found a prevalence of reliance on intermediaries, with the authors additionally summarising current UK and EU regulatory challenges \ie `know your customer'; anti-money laundering; and data protection. There are additional challenges of not only supporting the demonstration of the technology's unique trust-supporting benefits but of communicating DP data in a visual form that supports intelligibility amongst non-expert users.  TAPESTRY addresses this challenge through designing visualisations of trust evidence collected from social media, that communicate the coherence (and by extension, provenance) of the digital personhood without disclosing specifics about a subject's past activities.  Related to TAPESTRY's ability to evidence social media activity in privacy preserving manner are distributed privacy-preserving social networks, such as Safebook \cite{CutilloMS09-Safebook} and early attempts to realize the functionalities of a social network using cryptographic techniques \cite{GuntherMS11,BarthBW06}. The focus of these approaches have been on the privacy of the users and allowing them full control over their data. Thus, they allow the user to alter their data after previously committing to it. By contrast, TAPESTRY provides an immutable record of past social media activity that a user may share to evidence the provenance of their digital identity.  The use of Blockchain to provide such a service -- analogous to a de-centralised credit reference for identity --  is unique to TAPESTRY and addresses emerging concerns among the general public around the risks to privacy and security of siloing data within centralised services or organisations.

\section{Overview of the TAPESTRY Service}\label{sec:platform}
\bgroup\obeylines
\begin{figure}
    \centering
    \includegraphics[width=\linewidth]{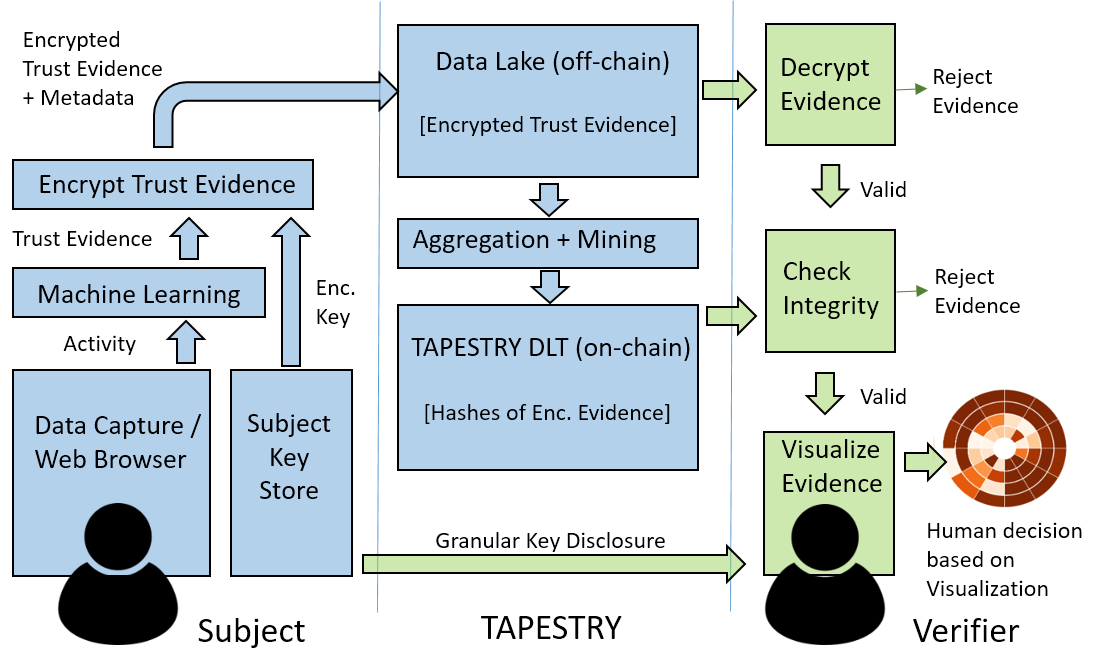}
    \caption{TAPESTRY System architecture.  Data is collected on subjects via the blue path. Digital activities (such as social media interactions) are captured on opt-in basis via a web browser extension.  Trust evidence is derived from those activities via a deep neural network (DNN) and encrypted using a secret key.  Keys differ between activities and change over time. Encrypted evidence is stored within a data lake, alongside metadata identifying the owner, timestamp and type of activity.  A hash of that metadata serves as a unique ID to that evidence.  The encrypted evidence is hashed alongside its unique ID within a proof-of-work blockchain.  The green path enables a verifier to check the provenance of a subject by requesting disclosure of the relevant decryption keys.  Encrypted evidence is requested from the lake; the provenance of that evidence is checked via the blockchain and it is decrypted.  Analysis of the DNN signals yields a visualisation gisting the relevant period of activity that helps the user make a trust decision on the subject.}
    \label{fig:storeflow}
\end{figure}

The TAPESTRY service collects, on an opt-in basis, signals derived from the  digital activities of a user (the 'subject') and enables that subject to securely share those signals with another user (the `verifier') in order to demonstrate their provenance of that subject's identity.  We assume that these signals or `trust evidence' (TE) are collected over longitudinal time periods from a rich tapestry of activities such as social media posts on various social platforms.  

In a typical interaction, the verifier will determine what kinds of TE are sufficient to make a human decision on the trustworthiness of a user according to their use context.  For example, on an online dating forum, or a ride-sharing service, a verifier might request evidence of a year of posting activity on any of several social media platforms in order to make their trust decision.  It is a matter for the subject to decide whether to disclose the requested evidence, and indeed the act of declining creates in itself a basis for  the verifying party to make a trust decision.

\subsection{Privacy Attributes}

In order to provide privacy to TAPESTRY users, TE is derived through a one-way hashing function that creates a compact, privacy-preserving gist of the semantic content of an activity (for example the text or image posted).  TAPESTRY utilises a deep neural network (DNN) to perform this distillation in order to prevent content from being recovered from TE, yet enabling two pieces of TE to be compared to quantify the similarity of the content that generated it. The details of this process are described further within Sec.~\ref{sec:ml}.  Thus a subject may share evidence of an activity, such as a social media post, without providing the content of that post to the verifier.  Furthermore, TE is stored within the platform in an encrypted form using a secret key held by the subject.  A different key is used for each TE generated from each type of activity (Facebook photo post, Twitter text post) and is changed periodically.  When a subject agrees to disclose TE to a verifier they do so by sharing the relevant keys.  Key generation and sharing, as well as the broader encryption scheme within TAPESTRY, is discussed in Sec.~\ref{sec:crypto}.

Since the volume of TE (\eg spanning months or years) requested of a subject is typically large, TAPESTRY creates a visual gist ('visualisation') of the TE in order to make it intuitively comprehensible to the verifier.  The design of the visualisation is discussed in Sec.~\ref{sec:viz}.  The core information communicated via the visualisation is the coherence of the user's digital history, derived from the timestamps and similarities of the TE shared by the subject.  Having sight of this visualisation the verifier finally makes a human decision as to whether the user is trustworthy, in combination with other external factors such as  social norms prevailing in their use context.  At no point is an automated decision offered to the verifier as to the trustworthiness of the user, and at no point is the subject's decision to share TE made automatically.  Rather, TAPESTRY acts as a privacy preserving conduit for the request and supply of TE.  

\subsection{De-centralised Trust Model}

TAPESTRY is designed around a decentralised trust model, without reliance upon third-parties to vouch for the integrity and provenance of TE.  This trust model is facilitated via a proof-of-work (PoW) Blockchain.

Recent changes in legislation (GDPR \cite{GDPR}) mean that users have the right to request that their personal data be deleted from any systems controlled by third-parties. In the case of TAPESTRY this meant that raw data from user activities could not be stored on-chain, as this could not be later removed. Storing only the encrypted vectors from the machine learning models, would also not comply due to the nature of the computations involved, which create an alternative digital representation of the user's behavior, thus personally identifiable information. Storing the personally identifiable information off-chain, within one or more independent data lakes (DLs), provides a method of data capture that facilitates recovering data for verification purposes, and can be deleted at the user's request.

TAPESTRY therefore uses a hybrid system of on- and off-chain storage for TE; see Fig.~\ref{fig:storeflow}.  A cloud service (of which many independently operated services are assumed to exist) maintains the DL into which the subject commits encrypted TE alongside plaintext metadata that identifies the user uniquely, as well as the timestamp and type of activity.  A SHA-256 hash of the encrypted TE is stored within a proof-of-work (PoW) Blockchain, keyed by a hash of the metadata (computed also via SHA-256) which serves as a unique identifier to the TE record in the DL.  In practice, a block committed to the PoW chain contains many such pairs.  Fig.~\ref{fig:actstorage} summarises the interactions between the subject, the DL and the Blockchain during collection of TE.  

On-chain hashing enables the verifier to check the provenance of TE, prior to decrypting and visualizing that evidence for human judgement.  The hash of the encrypted evidence received is compared to that stored immutably within the Blockchain.  This guards against an attack via fabrication of TE by the DL provider.  The PoW Blockchain is implemented via Ethereum, and a  smart contract to fetch (\ie search and retrieve) and to commit (append) data to the Blockchain is provided.  Failure to verify the provenance of the data, or to decrypt to the data into a parseable form (\eg due to an invalid secret key supplied by the subject) results in an immediate rejection of the TE and strongly implies an untrustworthy interaction.  Fig.~\ref{fig:actverify} summarises the interactions between the verifier, subject, DL (one pictured) and the public Blockchain during sharing and verification of TE.

Given the complex security model employed to ensure secure data transmission and storage, it is plausible that user error could cause the loss of the keys, rendering their TAPESTRY data inaccessible. Users could wish to share their keys with a trusted third party key store, which could provide their keys when required or even act as a facilitator during the verification process.  This optional step is analogous to sharing the private keys of cryptocurrency wallets with a centralised brokering service.

\begin{figure}
\centering
\hspace*{-2cm}\begin{minipage}{0.8\linewidth}
\centering
\def\Client{Subject}
\def\ClientMid{ClientMid}
\def\ClientDataLake{ClientDataLake}
\def\DataLake{Data Lake}
\def\DataLakeDLT{DataLakeDLT}
\def\DataLakeDLTMid{DataLakeDLTMid}
\def\DLTMid{DLTMid}
\def\DLT{DLT}
\begin{tikzpicture}[every node/.style={font=\normalsize,minimum height=0.5cm,minimum width=0.5cm}]
\node[matrix, very thin,column sep=0.2cm,row sep=0.25cm] (matrix) at (0,0){
  & \node(0,0) (\Client) {}; &&&&&&&& 
  \node(0,0) (\DataLake) {}; &&&&&&&&
  \node(0,0) (\DLT) {}; & \\
  & \node(0,0) (\Client 1) {}; &&
  \node(0,0) (\ClientMid 1) {}; &&&
  \node(0,0) (\ClientDataLake 1) {}; &&&&&&&&&&&& \\
  & \node(0,0) (\Client 2) {}; &&
  \node(0,0) (\ClientMid 2) {}; &&
  \node(0,0) (\ClientDataLake 2) {}; &&&&&&&&&&&&& \\
  & \node(0,0) (\Client 3) {}; &&
  \node(0,0) (\ClientMid 3) {}; &&&
  \node(0,0) (\ClientDataLake 3) {}; &&&&&&&&&&&& \\
  & \node(0,0) (\Client 4) {}; &&
  \node(0,0) (\ClientMid 4) {}; &&
  \node(0,0) (\ClientDataLake 4) {}; &&&&&&&&&&&&& \\
  & \node(0,0) (\Client 5) {}; &&
  \node(0,0) (\ClientMid 5) {}; &&&
  \node(0,0) (\ClientDataLake 5) {}; &&&&&&&&&&&& \\
  & \node(0,0) (\Client 6) {}; &&
  \node(0,0) (\ClientMid 6) {}; &&
  \node(0,0) (\ClientDataLake 6) {}; &&&&&&&&&&&&& \\
  & \node(0,0) (\Client 7) {}; &&&&
  \node(0,0) (\ClientDataLake 7) {}; &&&&
  \node(0,0) (\DataLake 7) {}; &&&&&&&&
  \node(0,0) (\DLT 7) {}; & \\
  & \node(0,0) (\Client 8) {}; &&&&&&&&
  \node(0,0) (\DataLake 8) {}; &&&&&
  \node(0,0) (\DataLakeDLT 8) {}; &&&
  \node(0,0) (\DLT 8) {}; & \\
  & \node(0,0) (\Client 9) {}; &&&&&&&&
  \node(0,0) (\DataLake 9) {}; &&&&&&
  \node(0,0) (\DLTMid 9) {}; &&
  \node(0,0) (\DLT 9) {}; & \\
  & \node(0,0) (\Client 10) {}; &&&&&&&&
  \node(0,0) (\DataLake 10) {}; &&&&
  \node(0,0) (\DataLakeDLT 10){};&& 
  \node(0,0) (\DLTMid 10) {};&&
  \node(0,0) (\DLT 10) {}; & \\
  & \node(0,0) (\Client 11) {}; &&&&&&&&
  \node(0,0) (\DataLake 11) {}; &&&&&
  \node(0,0) (\DataLakeDLT 11){};&&&
  \node(0,0) (\DLT 11) {}; & \\
  & \node(0,0) (\Client 12) {}; &&&&&&&&
  \node(0,0) (\DataLake 12) {}; &&&&&&&&
  \node(0,0) (\DLT 12) {}; & \\};
\fill 
	(\Client) node[draw,fill=white] {\Client}
	(\DataLake) node[draw,fill=white] {\DataLake}
	(\DLT) node[draw,fill=white] {\DLT};
\draw [dashed] 
  (\Client) -- (\Client 12)
  (\DataLake) -- (\DataLake 12)
  (\DLT) -- (\DLT 12);
\filldraw[fill=gray!20]
  (\Client 1.north west) rectangle (\Client 7.south east)
  (\DataLake 7.north west) rectangle (\DataLake 11.south east)
  (\DLT 8.north west) rectangle (\DLT 11.south east);
\draw [-latex] (\Client 1) -- (\ClientMid 1.east)-- (\ClientMid 2.east) -- (\Client 2);
\draw [-latex] (\Client 3) -- (\ClientMid 3.east)-- (\ClientMid 4.east) -- (\Client 4);
\draw [-latex] (\Client 5) -- (\ClientMid 5.east)-- (\ClientMid 6.east) -- (\Client 6);
\draw [-latex] (\Client 7) -- (\DataLake 7);
\draw [-latex] (\DataLake 8) -- (\DLT 8);
\draw [-latex] (\DLT 9) -- (\DLTMid 9.west)-- (\DLTMid 10.west) -- (\DLT 10);
\draw [-latex] (\DLT 11) -- (\DataLake 11);

\fill
  (\ClientDataLake 1) node[below,text width=2cm,align=center] {Receive user activity}
  (\ClientDataLake 3) node[below,text width=2cm,align=center] {Calculate vector}
  (\ClientDataLake 5) node[below,text width=3cm,align=center] {Encrypt and sign vector}
  (\ClientDataLake 7) node[below] {Submit data}
  (\DataLakeDLT 8) node[above,text width=2cm,align=center] {Submit smart contract transaction}
  (\DataLakeDLT 10) node[above,text width=2cm,align=center] {Transaction stored in new block}
  (\DataLakeDLT 11) node[below] {Transaction hash};
\end{tikzpicture}
\end{minipage}
    \caption{Sequence diagram of the collection of trust evidence (TE) from a subject. TE is generated locally, in the form of a vector distilled from raw content via a deep neural network (DNN).  The vector is encrypted, then sent to the data lake (DL) which stores the encrypted TE off-chain and records a hash of it within a new block in the PoW Blockchain (DLT).}
    \label{fig:actstorage}
\end{figure}
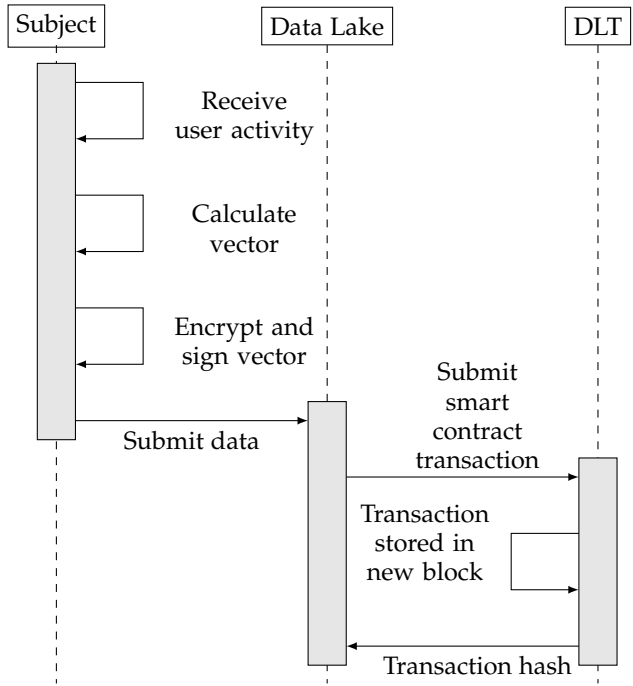

\begin{figure}
\centering
\hspace*{-2cm}\begin{minipage}{0.8\linewidth}
\centering
\def\Client{Subject}
\def\ClientVerifierMid{ClientVerifierMid}
\def\ClientVVMid{ClientVVMid}
\def\VerifierClientMid{VerifierClientMid}
\def\Verifier{Verifier}
\def\VerifierMid{VerifierMid}
\def\VerifierDataLakeMid{VerifierDataLakeMid}
\def\DataLake{Data Lake}
\def\DLT{DLT}
\begin{tikzpicture}[every node/.style={font=\normalsize,minimum height=0.5cm,minimum width=0.475cm}]
\node[matrix, very thin,column sep=0.15cm,row sep=0.3cm] (matrix) at (0,0){
& \node(0,0) (\Client) {}; &&&&& 
    \node(0,0) (\Verifier) {}; &&&&&&&&
    \node(0,0) (\DataLake) {}; &&&&&&
    \node(0,0) (\DLT) {}; & \\
& \node(0,0) (\Client 1) {}; &&&
    \node(0,0) (\ClientVerifierMid 1) {}; &&
    \node(0,0) (\Verifier 1) {}; &&&&&&&&
    \node(0,0) (\DataLake 1) {}; &&&&&&
    \node(0,0) (\DLT 1) {}; & \\
& \node(0,0) (\Client 2) {}; &&&&&
    \node(0,0) (\Verifier 2) {}; &&&&
    \node(0,0) (\VerifierDataLakeMid 2) {}; &&&&
    \node(0,0) (\DataLake 2) {}; &&&&&&
    \node(0,0) (\DLT 2) {}; & \\
& \node(0,0) (\Client 3) {}; &&&&&
    \node(0,0) (\Verifier 3) {}; &&&&
    \node(0,0) (\VerifierDataLakeMid 3) {}; &&&&
    \node(0,0) (\DataLake 3) {}; &&&&&&&
    \node(0,0) (\DLT 3) {}; & \\
& \node(0,0) (\Client 4) {}; &&
    \node(0,0) (\ClientVerifierMid 4) {}; &&
    \node(0,0) (\ClientVVMid 4) {}; 
    \node(0,0) (\Verifier 4) {}; &&&&&&&&&
    \node(0,0) (\DataLake 4) {}; &&&&&&&
    \node(0,0) (\DLT 4) {}; & \\
& \node(0,0) (\Client 5) {}; &&
    \node(0,0) (\ClientVerifierMid 5) {};&&
    \node(0,0) (\ClientVVMid 5) {}; &
    \node(0,0) (\Verifier 5) {}; &&&
    \node(0,0) (\VerifierMid 5) {}; &
    \node(0,0) (\VerifierDataLakeMid 5) {}; &&&&
    \node(0,0) (\DataLake 5) {}; &&&&&&
    \node(0,0) (\DLT 5) {}; & \\
& \node(0,0) (\Client 6) {}; &&&&&
    \node(0,0) (\Verifier 6) {}; &&&
    \node(0,0) (\VerifierMid 6) {}; &
    \node(0,0) (\VerifierDataLakeMid 6) {}; &&&&
    \node(0,0) (\DataLake 6) {}; &&&&&&
    \node(0,0) (\DLT 6) {}; & \\
& \node(0,0) (\Client 7) {}; &&&&& 
    \node(0,0) (\Verifier 7) {}; &&&
    \node(0,0) (\VerifierMid 7) {}; &
    \node(0,0) (\VerifierDataLakeMid 7) {}; &&&&
    \node(0,0) (\DataLake 7) {}; &&&&&&
    \node(0,0) (\DLT 7) {}; & \\
& \node(0,0) (\Client 8) {}; &&&&& 
    \node(0,0) (\Verifier 8) {}; &&&
    \node(0,0) (\VerifierMid 8) {}; &&&&&
    \node(0,0) (\DataLake 8) {}; &&&&&&
    \node(0,0) (\DLT 8) {}; & \\
& \node(0,0) (\Client 9) {}; &&&&& 
    \node(0,0) (\Verifier 9) {}; &&&&&
    \node(0,0) (\VerifierMid 9) {}; &&&
    \node(0,0) (\DataLake 9) {}; &&&&&&
    \node(0,0) (\DLT 9) {}; & \\
& \node(0,0) (\Client 10) {}; &&&&& 
    \node(0,0) (\Verifier 10) {}; &&&&&
    \node(0,0) (\VerifierMid 10) {}; &&&
    \node(0,0) (\DataLake 10) {}; &&&&&&
    \node(0,0) (\DLT 10) {}; & \\
& \node(0,0) (\Client 11) {}; &&&&&
    \node(0,0) (\Verifier 11) {}; &&&
    \node(0,0) (\VerifierMid 11) {}; &&&&&
    \node(0,0) (\DataLake 11) {}; &&&&&&
    \node(0,0) (\DLT 11) {}; & \\
& \node(0,0) (\Client 12) {}; &&&&&
    \node(0,0) (\Verifier 12) {}; &&&
    \node(0,0) (\VerifierMid 12) {}; &&&&&
    \node(0,0) (\DataLake 12) {}; &&&&&&
    \node(0,0) (\DLT 12) {}; & \\
& \node(0,0) (\Client 13) {}; &&&&&
    \node(0,0) (\Verifier 13) {}; &&&
    \node(0,0) (\VerifierMid 13) {}; &&&&&
    \node(0,0) (\DataLake 13) {}; &&&&&&
    \node(0,0) (\DLT 13) {}; & \\
& \node(0,0) (\Client 14) {}; &&&&&
    \node(0,0) (\Verifier 14) {}; &&&
    \node(0,0) (\VerifierMid 14) {}; &&&&&
    \node(0,0) (\DataLake 14) {}; &&&&&&
    \node(0,0) (\DLT 14) {}; & \\
& \node(0,0) (\Client 15) {}; &&&&&
    \node(0,0) (\Verifier 15) {}; &&&
    \node(0,0) (\VerifierMid 15) {}; &&&&&
    \node(0,0) (\DataLake 15) {}; &&&&&&
    \node(0,0) (\DLT 15) {}; & \\
& \node(0,0) (\Client 16) {}; &&&
    \node(0,0) (\ClientVerifierMid 16) {}; && 
    \node(0,0) (\Verifier 16) {}; &&&
    \node(0,0) (\VerifierMid 16) {}; &&&&&
    \node(0,0) (\DataLake 16) {}; &&&&&&
    \node(0,0) (\DLT 16) {}; & \\
& \node(0,0) (\Client 17) {}; &&
    \node(0,0) (\ClientVerifierMid 17) {}; &&&
    \node(0,0) (\Verifier 17) {}; &&
    \node(0,0) (\VerifierMid 17) {}; &&&&&&
    \node(0,0) (\DataLake 17) {}; &&&&&&&
    \node(0,0) (\DLT 17) {}; & \\
& \node(0,0) (\Client 20) {}; &&&&&
    \node(0,0) (\Verifier 20) {}; &&&&&&&&
    \node(0,0) (\DataLake 20) {}; &&&&&&
    \node(0,0) (\DLT 20) {}; & \\};
\fill 
    (\Client) node[draw,fill=white] {\Client}
    (\Verifier) node[draw,fill=white] {\Verifier}
    (\DataLake) node[draw,fill=white] {\DataLake}
    (\DLT) node[draw,fill=white] {\DLT};
\draw [dashed] 
    (\Client) -- (\Client 20)
    (\Verifier) -- (\Verifier 20)
    (\DataLake) -- (\DataLake 20)
    (\DLT) -- (\DLT 20);
\filldraw[fill=gray!20]
    (\Client 1.north west) rectangle (\Client 1.south east)
    (\Verifier 1.north west) rectangle (\Verifier 16.south east)
    (\DataLake 2.north west) rectangle (\DataLake 3.south east)
    (\DLT 9.north west) rectangle (\DLT 10.south east);
\draw [-latex] (\Client 1) -- (\Verifier 1);
\draw [-latex] (\Verifier 2) -- (\DataLake 2);
\draw [-latex] (\DataLake 3) -- (\Verifier 3);
\draw [-latex] (\Verifier 5) -- (\VerifierMid 5.west) -- (\VerifierMid 6.west) -- (\Verifier 6);
\draw [-latex] (\Verifier 7) -- (\VerifierMid 7.west) -- (\VerifierMid 8.west) -- (\Verifier 8);
\draw [-latex] (\Verifier 9) -- (\DLT 9);
\draw [-latex] (\DLT 10) -- (\Verifier 10);
\draw [-latex] (\Verifier 11) -- (\VerifierMid 11.west) -- (\VerifierMid 12.west) -- (\Verifier 12);
\draw [-latex] (\Verifier 13) -- (\VerifierMid 13.west) -- (\VerifierMid 14.west) -- (\Verifier 14);
\draw [-latex] (\Verifier 15) -- (\VerifierMid 15.west) -- (\VerifierMid 16.west) -- (\Verifier 16);

\draw [draw=black, fill opacity=0.2] (\ClientVerifierMid 4.north west) -- (\DLT 4.north west) -- (\DLT 17.south west) -- (\ClientVerifierMid 17.south west) -- cycle;
\draw [line width=0.3mm, draw=black, fill opacity=0.2] (\ClientVerifierMid 4.north west) -- (\ClientVVMid 4.north east) -- (\ClientVVMid 5.north east)  -- (\ClientVVMid 5.south west) -- (\ClientVerifierMid 5.south west) -- cycle;

\fill
(\ClientVerifierMid 1) node[below,text width=2cm,align=left] {Share keys for date range}
(\VerifierDataLakeMid 2) node[above,text width=3cm,align=left] {Request activities}
(\VerifierDataLakeMid 3) node[above,text width=3cm,align=left] {Return  $n$ activities}
(\VerifierMid 5) node[above,text width=2cm,align=center] {Validate signatures}
(\VerifierMid 7) node[right,text width=2cm,align=left] {Calculate SHA256 hashes}
(\VerifierMid 9) node[above,text width=4cm,align=left] {Request hash for activity}
(\VerifierMid 10) node[above,text width=4cm,align=left] {Return hash}
(\VerifierMid 11) node[right,text width=1.5cm,align=left] {Compare hashes}
(\VerifierMid 13) node[right,text width=1.5cm,align=left] {Decrypt vectors}
(\VerifierMid 15) node[right,text width=2cm,align=left] {Produce visualisation}
(\ClientVerifierMid 4) node[below right,text width=1.2cm,align=center] {loop $n$ times};
\end{tikzpicture}
\end{minipage}
\caption{Sequence diagram of the process by which a verifier determines the trustworthiness of a subject. The subject shares relevant secret keys with the verifier, enabling encrypted TE vectors to be retrieved from the data lake, verified for provenance against the Blockchain (DLT), and decrypted.  The vectors are converted into a visualisation to aid the verifier's trust decision.}
\label{fig:actverify}
\end{figure}
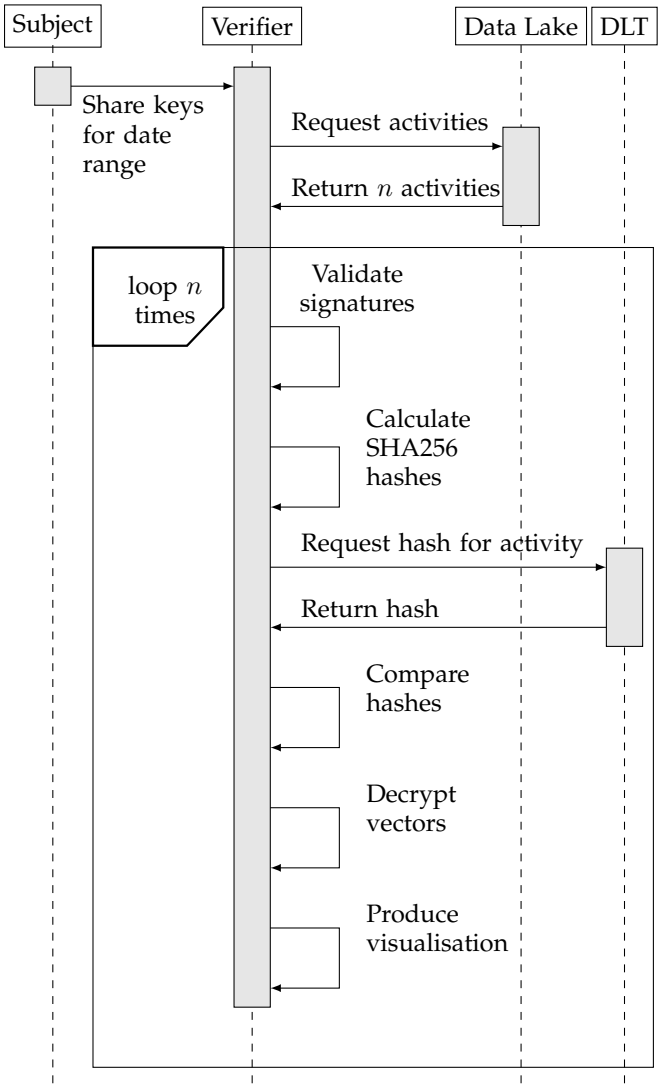
\egroup
\section{Extracting Trust Evidence}
\label{sec:ml}




We now describe the process through which activity content is hashed into trust evidence (TE).  Our approach is based on the hypothesis that people have consistent (or slowly evolving) behavior and personal interests over longitudinal time periods~\cite{viswanath2014towards,wang2018composite}. Deviations from this organic behavior pattern indicate either a non-natural (fake) account \eg as a vehicle for spam or online scam, or a legitimate account that has been hijacked for similar purpose resulting in abnormal characteristics in the timeline.

Therefore, we tackle this deviation detection problem from behavioral norm as an outlier/anomaly detection task.  Our goal is therefore to reduce the content of a post to a compact, real-valued vector (the TE) such that similar semantic content maps to similar TE. In this work, we present our deep neural network (DNN) based method to detect coherent/incoherent activities on the Twitter social media platform via analysis of text within a subject's posts.  Through a combination of semantic embedding and temporal modelling, we map activity content to TE and leverage the  temporal coherence of TE to help prove the provenance of a digital identity.

\subsection{Data pre-processing} 
User-generated content (UGC) on social media is often a mixture of texts, special characters, hashtags, emojis and links. This kind of raw data is not directly suitable for machine learning method. In natural language processing, a pre-processing step is necessary to clean the data to normalize it for the learning process. In this paper, we consider only meaningful texts and focus on topic analysis. We first remove special characters and retain only texts in a tweet. We then apply successive operations, including tokenisation, stop words removal and stemming, and lemmatisation.

\subsection {Topic Word Modeling}
A word embedding is a distributed representation of words, incorporating semantic information \cite{mikolov2013distributed} that is learned from a large corpus of text (all tweets in the collected data set in our case). Topical modeling \cite{blei2003latent} extracts a distribution of words as topics, and a distribution of topics as documents. We implemented topic word embeddings, as proposed in \cite{liu2015topical}, to capture contextual information in the given document. A topic word embedding is considered as a word-topic pair $< w_i, t_i >$. We considered all the tweets from one user as a document. The learned feature can enhance discrimination between words in different contexts and styles. A tweet embedding is the average of all topic word embeddings derived from the words in the tweet. 

\subsection{Temporal Coherence via Long-Short Term Memory}
The application of Deep Learning~\cite{goodfellow2016deep} is proving highly effective in making sense of signals in computer vision~\cite{krizhevsky2012imagenet}, natural language~\cite{goldberg2016primer} and robotics~\cite{mnih2015human}. We apply a DNN to learn features for each user in a temporal window, denoted as user embedding. The user embedding is regarded as a temporal pattern of tweets in a fixed time window.  
 
The Long-Short Term Memory (LSTM) model~\cite{hochreiter1997long} is a recurrent neural network used to model and predict time-series data. We built a sequence model to capture the coherence activities using a LSTM model and trained to extract the user's behavior norm based on their `daily story', e.g. as played out on social media or through other online activity. We implemented a bi-directional LSTM to model the temporal coherence on a daily and weekly basis across the captured Twitter data (temporal segment). We adopted a two-layer bidirectional LSTM, followed by two fully connected layers. The input of LSTM is the topic word embedding and the output is a daily or weekly tweets embedding. 

\subsection{Triplet network for TE Embedding}

In order to compare TE from activities over time, it is necessary to learn a metric embedding in which norms may be computed to quantify deviations in the topic word embedding over time.  

Triplet DNNs have been used more broadly to learn such embeddings for information retrieval \eg for visual search  \cite{CTU-ECCV2016,gordo2016deep} and we similarly perform supervised learning of the TE embedding using a triplet network strategy~\cite{schroff2015facenet}. The objective of this network structure is to map TE within the topic word embedding to a metric embedding in which similar TE samples are pushed together and dissimilar samples pushed away from each other in the learned space. Here, similar samples are the temporal segments from one individual and dissimilar samples are the ones from different individuals. The method  proved efficient in identifying different individuals from their temporal features, as learned by the prior LSTM step. We tested the method to detect compromised moments of an account, by randomly selecting a time step on one user's time-line feed. We then replaced the Tweets after the time point by the tweets from another user in order to simulate anomalous accounts, in order to provide negative exemplars for training. 

\begin{figure}[t!]
\centering
\includegraphics[width=\linewidth]{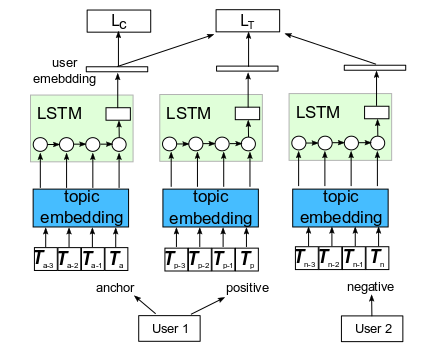}
    \caption{Triplet LSTM architecture of our DNN used to extract TE from activity content.  Content is converted to a topic embedding via averaging words within a tweet under a learned semantic (topic) model.  Temporal embedding is learned via LSTM reflecting social media behavior over time.  Initially the LSTM is trained as a classifier to discriminate between users within a training corpus (under cross entropy loss $L_C$, yielding a 'user' embedding)  The embedding is then fine-tuned to yield the final TE embedding via triplet training using  sequences of real twitter posts (positives), and simulated fake sequences created by splicing twitter histories together negatives), under a triplet loss $L_T$. }~\label{fig:network}
\end{figure}

Fig.~\ref{fig:network} illustrates our network architecture and end-to-end pipeline of our machine learning algorithm for TE extraction. Given a set of $n$ users $\mathcal{U} = \{u_1, u_2, \cdots, u_n\}$, each user has a sequence of $k$ tweets $u_i = \{T_1, T_2, \cdots, T_k\}$. The tweets are first pre-processed and projected from variable length text strings to the topic embedding space. The LSTM then learns a temporal feature of the tweets for each temporal window, denoted as user embedding, $ue_i = \{twe_1, twe_2, \cdots, twe_{k-w+1}\}$, where $w$ is the window size. The learning strategy is a classification followed by a triplet network fine-tuning.  We use a single cross-entropy loss $\mathcal{L}_C$ for pre-training and the a combined classification and triplet loss $\mathcal{L}_T$ based on $L^2$ norm for fine-tuning \cite{schroff2015facenet}. In our experiments we use $n=8000$ and $k=800$ to train the TE embedding (c.f. Sec.~\ref{sec:hijack}).

\section{Privacy and Information Security}
\label{sec:crypto}

In this section we elaborate on the data encryption and key sharing features of the TAPESTRY platform, the security properties they require and the security guarantees they provide for the service. We follow the formalism of timeline activity proofs (TAP) \cite{DraganM18}, where we have the user's trust evidence as an activity.

\subsection{Entities and their roles}

\paragraph{\it Subjects} 
We model subjects \ie users on whom data is collected, using secret-public key pairs $(\sk,\pk)$, with $\sk$ used generically to contain all secret information required by the user, i.e. signing keys and seed for derivation of encryption key. The public key $\pk$ is used as public identifier for the user, referred to in Sec.~\ref{sec:platform} as the user's unique ID.  This is analogous to a public Bitcoin wallet address, and we allow the user to create multiple identities, and thus, hold multiple key pairs.  An assumption of our system is that a physical identity can not transfer ownership or operation of such key pairs, so grounding the key pair as a basis for identity in TAPESTRY.

\paragraph{\it Public Ledgers with external database} 
For simplicity, we consider an idealized version of the blockchain approach in Section \ref{sec:platform}, where the public ledger contains an {\em append-only list} and an {\em external database}. This allows us to introduce search functionalities that verifiers require, and offer data removal to the users that have submitted their activities.

\paragraph{\it Verifiers and Policies}  Verifiers establish {\em policies} - statements over different types of activities in specific intervals, that the users must satisfy. Our construction follows  \cite{DraganM18}, except we consider a particular type of policy where a human verifier makes a binary decision based on the visualizations of Sec.~\ref{sec:viz}.

\subsection{Form of Trust Evidence} 

A user's TE maintains a strict format: 
the user's public identity $\pk$ (\ie unique identified), the time it has been registered $\ltime$, the type of evidence $\atype$ with any optional descriptors $\atags$, the  machine learning encoding (\ie real-valued vector) of the evidence data $\adata$, and a digital signature $\signature$ to authenticate that it was submitted by user $\pk$. When the user submits this trust evidence, the data component is encrypted $\cdata$, therefore:
$$\mathrm{TE}=\left<\pk,\ltime,\atype,\cdata, \atags, \signature\right>.
$$

\paragraph{\it Building Blocks}
Our construction relies on {\em pseudo-random functions} (PRF) \cite{GoldreichGM86-PRF} and {\em digital signature} (DS) \cite{DiffieH76-DigitalSign} that are {\em existentially unforgeable under chosen message attacks} (EUF-CMA) \cite{GoldwasserMY83-unforgeability}. Additionally, we consider a {\em symmetric encryption scheme } (SE) with two security requirements: {\em indistinguishability under chosen plaintext attacks} (IND-CPA) and {\em wrong key detection} (WKD) \cite{CanettiKVW10-symmetricencrypt}.

\paragraph{\it Key Management}
The trust evidence data is encrypted with a symmetric encryption scheme, where the encryption/decryption keys play an important part of the policy verification. Our solution is to derive unique encryption keys for finite time periods and for each kind of activity; in practice this could be as granualar as a key per piece of TE. We realize this by assigning a random PRF seed $s$ to each user, when they join our system. For $\mathrm{TE}$, the encryption key $ek$ is build as:
$$ ek=\mbox{PRF}(s,\mbox{PRF}(s,\pk,\atime),\atype). $$
For a greater degree of granularity, we may consider counting the same type of trust evidence received at the same time duration: $ \mbox{PRF}(s,ek, \counting)$. Furthermore, this allows for a granular disclosure of encryption keys only for the trust evidence required by verifiers without compromising the security of the other trust evidence.

\subsection{Security Properties}
There are two security properties that our system satisfies: {\em data confidentiality} that ensures the privacy of trust evidence data after the user has submitted it to the external database associated with the ledger, and {\em authentication policy compliance} where verifiers are only convinced by users who actually satisfy verification policies. Formal definitions are available in \cite{DraganM18}.

\paragraph{\it Data Confidentiality}
Intuitively, this property ensures that no information is revealed concerning the trust evidence data the user is submitting, just by analyzing entries in the ledger. This property is modeled by using a probabilistic polynomial-time (PPT) adversary that is required to distinguish between two private activity encodings by seeing an entry in the database that corresponds to one of them. The entries differ only on the data component, while the public key, the time, type and tags are the same for both entries. 
Following the formalism of TAP, we use cryptographic primitives that satisfy the security requirements of TAP: IND-CPA for the symmetric encryption scheme, and pseudo-randomness for PRF. 

\paragraph{\it Authenticated Policy Compliance} 
This property ensures that no malicious user can impersonate an honest user or fake the existence of trust evidence in the database associated to the ledger, and convince an honest verifier that they are authorized and satisfy their policy. We model this using a PPT adversary that can submit entries to the external database of a public ledger and is successful if he can convince an honest verifier to accept the evidence, when one of the two conditions are satisfied: either the adversary impersonated an honest user, or he provided a successful proof for a policy he does not satisfy. 
Following the formalism of TAP, we use cryptographic primitives that satisfy the security requirements of TAP: WKD for the symmetric encryption scheme, EUF-CMA for the digital signature, and pseudo-randomness for PRF.

\subsection{Implementation details}
Our cryptographic primitives are instantiated using the implementation from the python library {\em pynacl}. Our PRF is instantiated with the BLAKE2b \cite{pynacl}. In \cite{AumassonNWW13-blacke2b}, it has been shown that BLAKE2b satisfies the pseudo-randomness property required by PRFs. Our DS uses the Ed25519 \cite{BernsteinDLSY12} implementation from \cite{pynacl} to instantiate the digital signature. Ed25519 offers existential unforgeability under chosen message attacks. Our SE is instantiated with the Salsa20 and Poly1305 MAC \cite{housley-rfc-chacha20-poly1305}. In \cite{Procter14a-digitalsignature} it has been shown that this construction satisfies IND-CPA. Moreover, the exact construction uses the technique from \cite{CanettiKVW10-symmetricencrypt}, and therefore also satisfies WKD.



\section{Making Trust Evidence  Comprehensible}
\label{sec:viz}

\begin{figure}[t!]
\centering
\includegraphics[width=\linewidth]{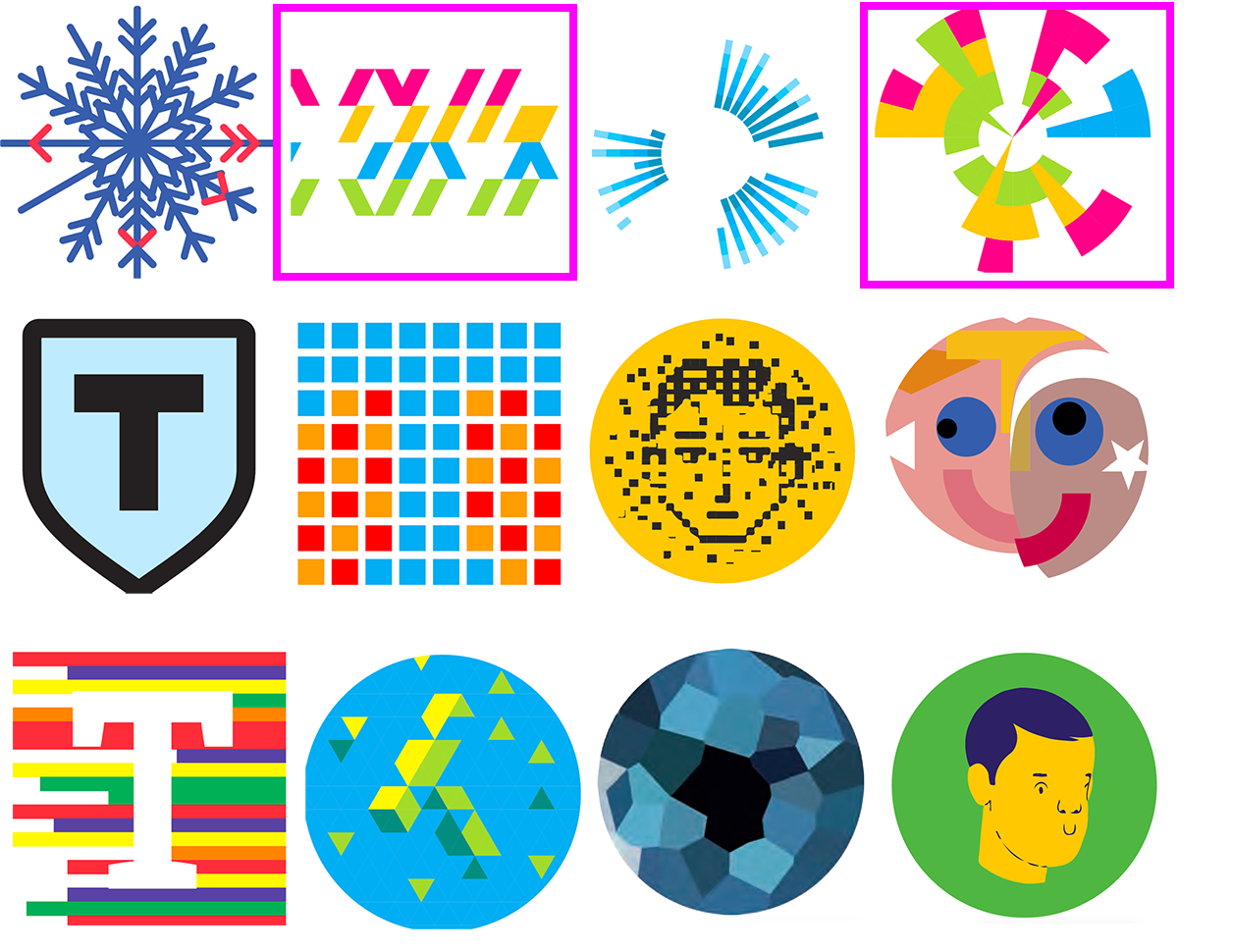}
    \caption{Initial designs for the TAPESTRY visualisation prototyped with the focus group.  The two designs progressed for evolution and implementation in the service, based on user feedback, were the `slash' and the `pie' (shown second and fourth from left on the top row).  The purpose of the visualisation is to communicate the completeness of TE records over the shared time period, and the coherence of activities generating that TE \ie to flag anomalous behaviour.  In many of the initial designs, these properties were reflected by spatial coverage and use of colour respectively. }~\label{fig:12-visualisations}
\end{figure}

TAPESTRY aims to visually communicate a gist of the completeness and coherence of a subject's TE over time, so that a verifier can make an informed choice as to whether to trust that subject.  We rejected the security related motifs (e.g. ticks, padlocks) that are often used in online systems to signal the efficacy of a particular security function or domain of use; e.g. proportionate red-amber-green traffic light scales as used for food packaging to indicate nutritional content ~\cite{Desai2018}; bronze, silver, gold hues often incorporated into badge, certificate or star rating symbolism. Such tropes convey trustworthiness as quantifiable and unequivocal (see ~\cite{Nurse2014}). All visualisations are persuasive to an extent ~\cite{Correll2019}; this has serious design implications as TAPESTRY does not (visually) verify an online actor's trustworthiness but aims to support individuals in making their own judgments about in whom and what to trust. 
\begin{figure*}[t!]
\centering
\includegraphics[width=0.45\linewidth,height=5cm]{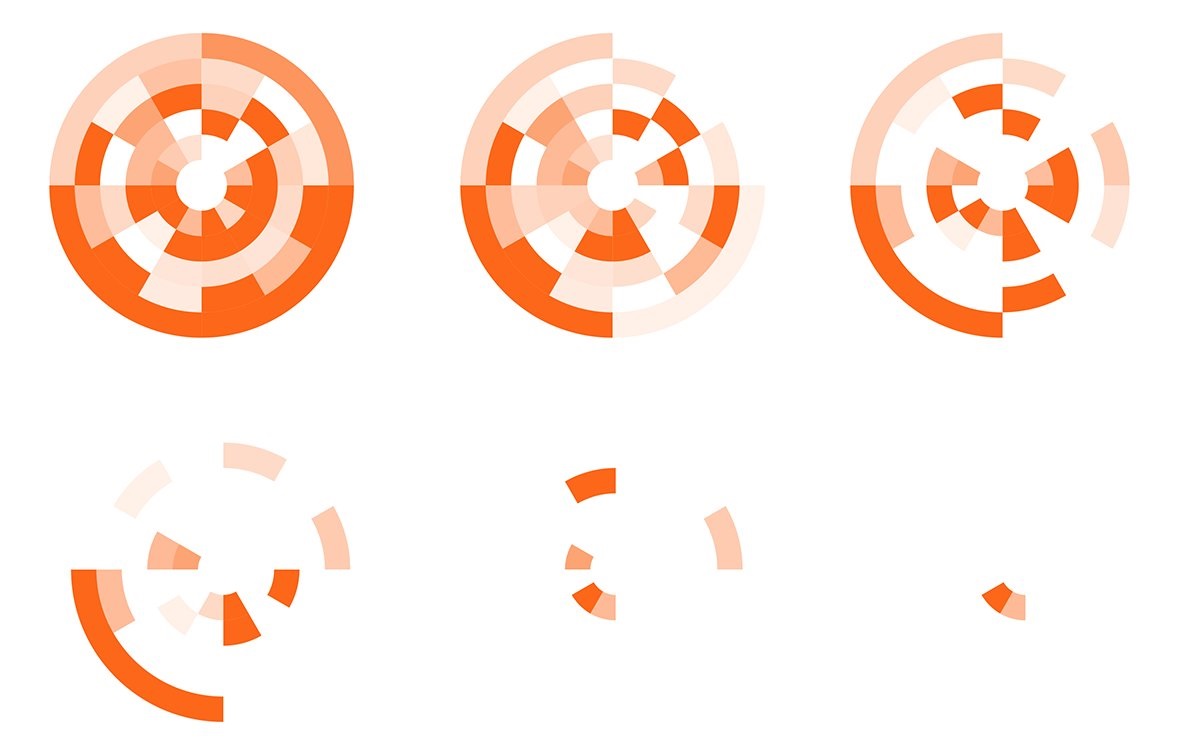}~~
    \includegraphics[width=0.45\linewidth,height=5cm]{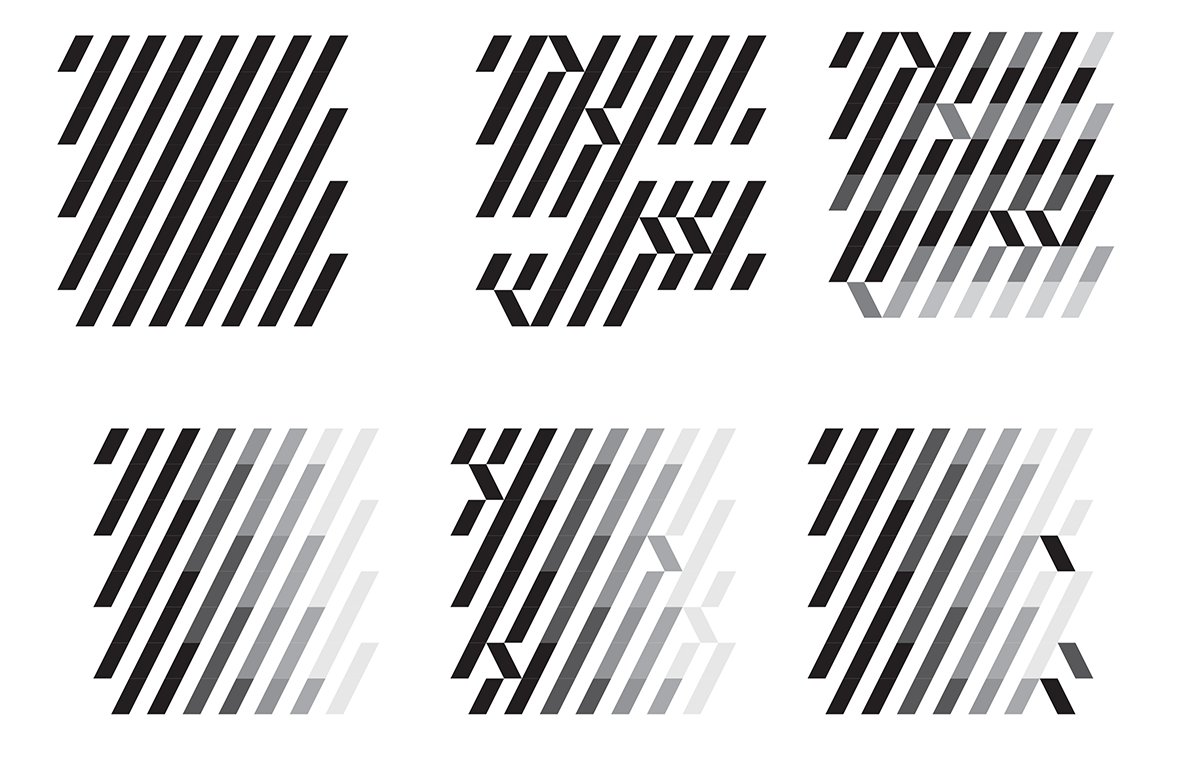}
    \caption{Developed visualisations deployed in TAPESTRY.  Left: Pie communicates interpersonal trust \eg to help users determine whether to trust an unknown business or service online.  Concentric rings of the pie correspond to different granularities of time, and shading is used to communicate coherence and volume of activity within the time period corresponding to each segment. Right: Slash  communicates introspective trust \eg to help users determine if their social media account has been compromised.  Each slash corresponds to a period of time, with backslashes indicating anomolous (outlier) TE during that period.  Shading is used to communicate volume of TE during that period, as with `pie'. Six instances of the visualisations are displayed ranging from complete and coherent, to sparse TE over the time period requested by the verifier.}~\label{fig:sm-evolution}
\end{figure*}

Key design challenges include integrating technically and conceptually complex models into human centred approaches ~\cite{Dove2017,Dove2014,Holmquist2017,Yang2018a,Yang2018b}, that account for different perspectives on human understandings, values and experience  ~\cite{Dourish2001, Elsden2018}.
TAPESTRY's target user group comprises a particular demographic that is less likely to use an online service's security features  ~\cite{Ofcom2018, Herley2009}; and has lower levels of numeracy (see ~\cite{Correll2019}), than on average. Thus, there are a multiplicity of interwoven design challenges and ethical concerns around visually or otherwise communicating TE and how this is then interpreted and understood, challenges that are amplified by the non-semantic nature of TE. 


\subsection{Prototyping of Visualisations}

User focus groups and lab-based workshops with user experience (UX) designers informed early concept designs for the visualisation. Our design inspirations were broad, from Knightmare, a 1990s British TV quest gameshow for children that manifested the health status of the characters using pixelated computer graphics, to more conventional information and data visualisation practices. From this we then produced 12 initial designs we called `snowflake'; `slash'; `radiance'; `pie'; `T-bar' (referring to TAPESTRY); `T'; `shield'; `tiles'; `pixel face'; `Picasso'; `eye' and `pixel head' (see Fig.~\ref{fig:12-visualisations}). We then rejected designs that could not depict sufficient granularity of either the completeness or coherence of TE over time.  We also rejected designs that did not readily scale down (i.e. for viewing on a small screen) \eg  `pixel head' inspired by Knightmare, the most anthropomorphic of the designs. We also rejected `eye' as evocative of a surveillance system. We explored use of colour and tone, both to enable additional granularity of visual representation of the shared TE and with regard to colour's culturally situated function that could invite potentially unintended meanings for some users. Additionally, in terms of interpreting completeness of TE -- within the research team it became apparent that the computer scientists  associated lighter tone with more TE while designers interpreted white areas of a design as an absence of TE within a given time period.  With these constraints in mind, we selected to use the idioms of `slash' and `pie' as the preliminary visualisations for further development.

\subsection{Evolved Visualisation Designs}

The final designs for the `pie' and `slash' TE visualisations are shown in Fig.~\ref{fig:sm-evolution}.  The choice of two designs for the visualisation reflect two use cases for deployment of TAPESTRY, evaluated in Sec.~\ref{sec:eval}.

Pie is based on a simple dial that lends itself to representing temporality and accumulation of DP over days, weeks, months etc, across concentric circles, as though accumulated TE is moving towards the core of the pie. This design is used for the interpersonal trust case in which users are required to make trust decisions on an {\em a priori} unknown online business or service.  We report of the efficacy of the visualisation in this context within the video games crowd-funding experiment of Sec.~\ref{sec:games}. 

Slash meanwhile was intended to communicate introspective trust, where users can check TE derived from their own digital footprint (\ie act as both subject and verifier) to determine whether their online accounts have been hacked (Sec.~\ref{sec:hijack}.). This required a design that could visually detail sudden dissonance within an otherwise relatively uniform pattern of DP as generated over time. A visual design analogy would be a ladder in hosiery or a dropped stitch in knitwear; these draw the eye, despite their small scale, to solicit a feeling of unease in the user to invite further investigation. 

Both these visualisations necessitate some initial explanation and guidance ~\cite{Correll2019}, though their intended meaning will require learning only once ~\cite{Cooper2009}. This is addressed via tutorial during the initial user sign-up to the TAPESTRY service. 


\begin{figure*}[t!]
\centering
\includegraphics[width=\textwidth]{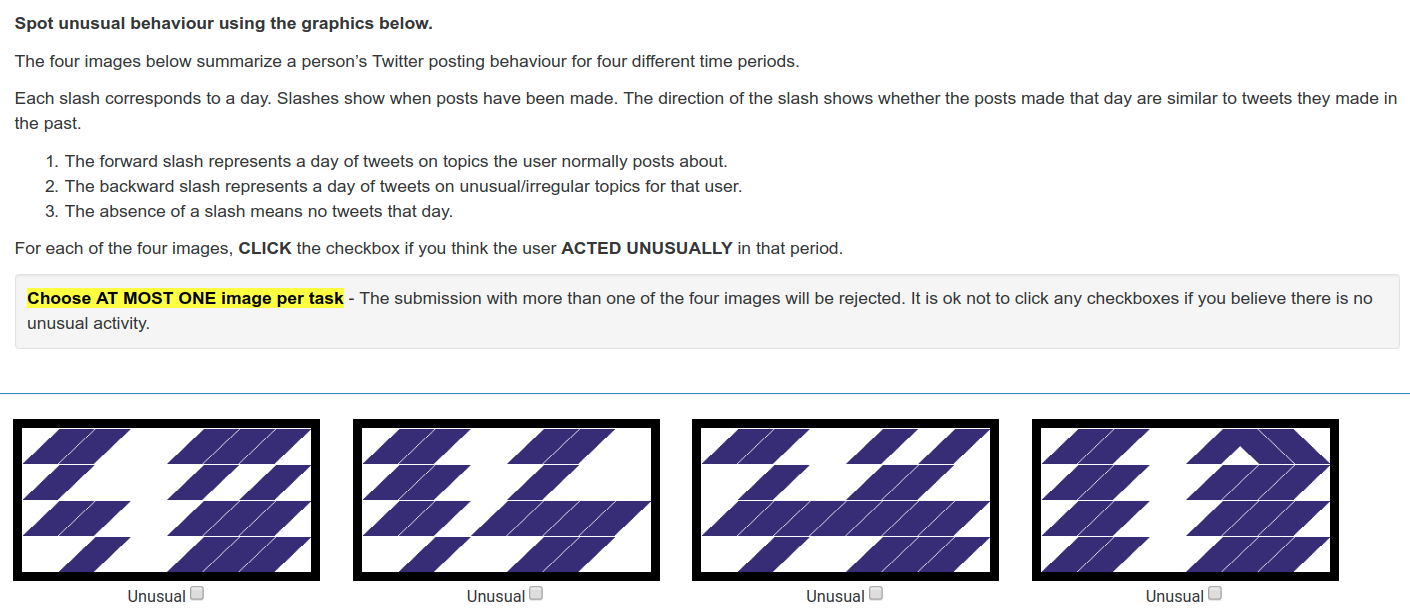}
    \caption{Mechanical Turk experiment evaluating the efficacy of the `slash' visualisation for anomaly detection (intraspective trust).  MTurk workers were presented with a series of four visualisations each summarizing four weeks of trust evidence.  Based on the series, users were asked which (if any) of the visualisations appeared anomalous relative to the others and so implied unusual activity was present in the social media feed.}~\label{fig:mturk}
\end{figure*}
\section{Experimental Evaluation}
\label{sec:eval}

We evaluate the TAPESTRY service in two contexts.  First, intraspective trust which we evaluate in the context of social media account hijacking; a user must determine whether their social account has been compromised due to anomalous posting behaviour.  Second, interpersonal trust in which a user must determine the trustworthiness of another online.  We  evaluate this in the domain of rewards-based crowd-funding, where parties are typically unknown to one another initially and a trust judgment is fundamental to the decision to invest in a venture.   

\subsection{Intraspective Trust: Anomaly Detection}
\label{sec:hijack}

We first evaluate the efficacy of our proposed DNN approach (Sec.~\ref{sec:ml}) for extracting trust evidence from social activities (text based Twitter posts). We evaluate its ability to discriminate between the behaviour of different users, and its ability to detect anomalies within the social media feed of individual users.  The experiments were conducted using a public dataset of Twitter posts (`tweets') gathered by Li et al.  ~\cite{li2012towards} initially comprising 50 million tweets for 140 thousand users. In our experiments, we study social media footprint over longitudinal time periods and clean the data by removing users with fewer than 800 tweets in their timeline feeds. The remaining 8000 users form the basis for our experiments.  

\subsubsection{Evaluating Trust Evidence (TE) Embedding}

\begin{figure}[t!]
\centering
\includegraphics[width=\linewidth]{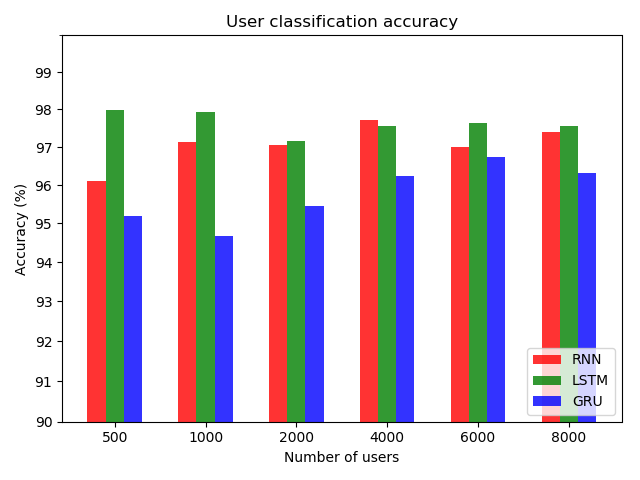}
    \caption{Evaluating user discrimination without our DNN learned embedding for trust evidence.  Our temporal modelling approach (based on LSTM) is compared to RNN and GRU sequence models.  The experiment is run for $n=[500,8000]$ corpus of users and accuracy measured as the $\% $ of users correctly identified based on 20\% of their TE.}~\label{fig:accuracy}
\end{figure}

We justify our choice of a LSTM to learn a temporal model for TE, via comparative evaluation against two state of the art DNN architectures; RNN and GRU.  We evaluate all three architectures as a user classification problem; the networks are trained using $80\%$ of the tweets of $n=[500,8000]$ users and tested on the remainder.  Accuracy is measured as the number of times the system correctly identifies the user among the $n$ possibilities.  Fig.~\ref{fig:accuracy} shows the result of classification accuracy on all the three models as $n$ increase. The LSTM model outperforms the other two models in most of cases.

\subsubsection{Evaluating Anomaly Detection}

We compare the efficacy of our learned embedding at detecting anomalies within a single user's history of TE.  For this experiment we train the model on 80\% of users, and testing on the remaining 20\%.  We compare several approaches to detecting anomalies within the test partition:


\begin{enumerate}
\item{{\bf One-class SVM (OCSVM)}~\cite{scholkopf2001estimating}  computes a non-linear boundary in a higher dimension space using kernel method for data project. This method allows for only positive data as 'one class'. }
\item{{\bf Isolation Forest (IF)}~\cite{liu2008isolation} evaluates the isolation degree of each data point using a random forest. This algorithm focuses on separating the outliers from the data points.}
\item{{\bf Local outliers factor (LOF)}~\cite{breunig2000lof} is a distance-based method using Euclidean distance considering the density of neighborhood information.}
\item{{\bf Proposed method.} We applied a distance based method outlier detection on our proposed user embedding. We evaluate Euclidean distance between the point and the average user embedding of first fraction of user's activity, represented as user behavior norm. We consider distance larger than a selected threshold as outliers.}
\end{enumerate}

We simulate hijacking actions on users' Twitter stream to evaluate anomaly detection. Given a fraction ($f_h$) hijacking length, we randomly replace fraction of tweets in each user's feed by another user's content. These replacements are unusual behaviors that deviate from user's activity norm. In Table~\ref{tab:ad}, we show the result of precision and recall. Our proposed method outperforms all the other methods and invariant to the number of fraction number of hijacking length.

\begin{table}[t!]
        \centering
        \begin{adjustbox}{width=.5\textwidth}
        \begin{tabular}{|c|c|c|c|c|c|c|c|c|c|}
        \hline
             & \multicolumn{3}{|c|}{Precision} & \multicolumn{3}{|c|}{Recall} & \multicolumn{3}{|c|}{F1-score}\\
            \hline
            $f_h$ & 0.1 & 0.2 & 0.3 & 0.1 & 0.2 & 0.3 & 0.1 & 0.2 & 0.3\\
            \hline
            \multicolumn{1}{|c|}{OCSVM} &0.43 & 0.46& 0.48 & 0.44 & 0.46 &0.48  & 0.43   & 0.46 & 0.47\\    
            \hline
            \multicolumn{1}{|c|}{IF} & 0.71 &0.62 & 0.57 & 0.75 & 0.64 & 0.59 & 0.73 &  0.63 & 0.58\\     
            \hline
            \multicolumn{1}{|c|}{LOF} & 0.12 &0.24 & 0.34 & 0.13 & 0.25 & 0.37 & 0.12 & 0.24 &0.35 \\    
            \hline
            \multicolumn{1}{|c|}{Ours} & \textbf{0.93} & \textbf{0.94}& \textbf{0.95} & \textbf{0.98} & \textbf{0.91} & \textbf{0.92} & \textbf{0.95} & \textbf{0.91} & \textbf{0.94}\\
            \hline
        \end{tabular}
        \end{adjustbox}
        \caption{Evaluating the ability of our embedding to perform anomaly detection.  Results compare our proposed approach to anomaly detection within the learned embedding to two common baselines. } 
        \label{tab:ad}
\end{table}

\subsubsection{Comprehensibility of the visualisation}
In order to evaluate if our proposed visualisation is understandable and helpful for non-expert users to detect potential anomalies, we  crowd-source evaluation on the Amazon Mechanical Turk (MTurk) platform. Fig.~\ref{fig:mturk} depicts a representative questionnaire. We provide four successive visualisations of of a user's activities sampled at regular intervals across their TE history. Each visualisation contains four weeks of activities. Normal behaviors are represented by forward slashes while the backward slashes as unusual behaviors detected by our proposed method in Section~\ref{sec:ml}. A blank position means no activities in that time-stamps. The MTurk worker is asked to select one if they think there exists unusual behaviors among the set of four visualisations \ie deviation from the norm. In the experiment, we collect 1000 users in our dataset and keep users who have over 4 weeks activities in their timeline, results 537 tasks. Each task is assigned to 5 different workers. A task is correctly detected if most of workers make the correct decision. As a result, 16 workers are attributed to the whole tasks and $93.8\%$ tasks were correctly selected by the workers. On average, each worker spend around 1 minute to finish a task including reading the instructions and submitting their decisions.

\subsection{Interpersonal Trust: Crowdfunding}
\label{sec:games}
We evaluate the performance of TAPESTRY in terms of benefit to the human decision making process when establishing the trustworthiness of an individual online. 

\subsubsection{Mock Crowd-funding Campaign}

We designed an interpersonal user study comprising a decision-making task experiment presented as a mock crowdfunding campaign in the context of the video games start-up; a sector heavily reliant upon such funding.  In this context, a user (the 'verifier') is invited to make an investment by another user (the 'subject') without prior knowledge of that subject.  Our game was offered for investment by eight crowdfunders on a mock platform; four real games industry professionals and four fake identities that had been created two months prior to the study.  We  commissioned an experienced video game narrative writer who mocked-up a pitch for new video game and crowdfunding campaign. We had gained the consent of the games developers to use their real profiles in the campaign. Meanwhile the writer produced four fake profiles based on their knowledge of the gaming industry. We created fake Twitter accounts for the fake profiles and continually tweeted relevant game and entrepreneur-related comments for two months prior to the study. All eight crowdfunders had one campaign web page hosted on a password protected micro-site, which included a description of the game (constant across all candidates), a short biography for each profile and a link to their Twitter account.  The creation date of the real and fake Twitter accounts was obfuscated.

\begin{figure}[t!]
\centering
    \includegraphics[width=\linewidth]{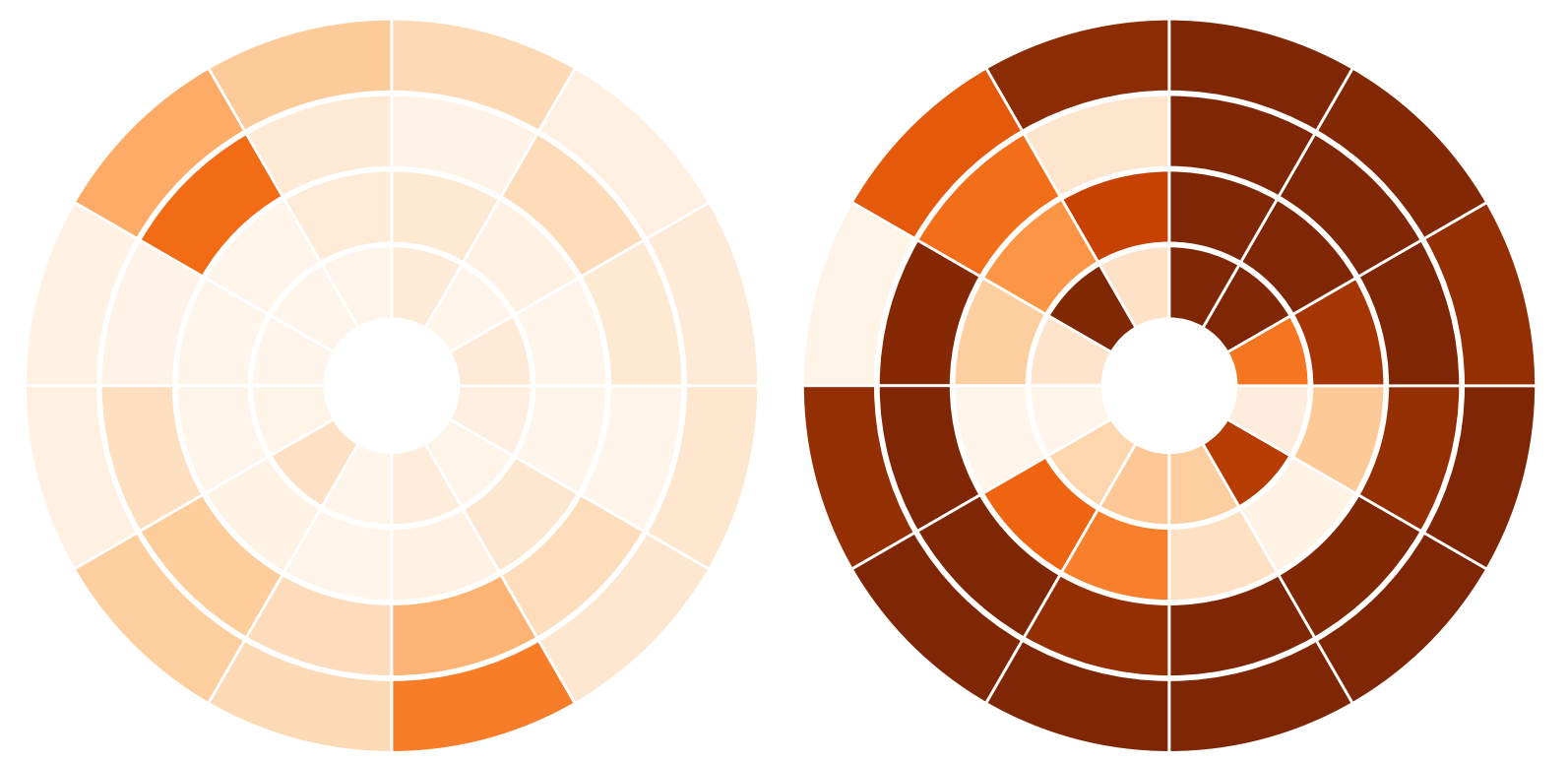}
    \caption{Visualisation of two crowdfunders' digital footprints: strong provenance (left) and limited provenance (right). The TE requested by the 'verifier' (participant) was fixed to 2 months with concentric circles represent daily tweet activity (inner) to weeks of activity (outer).}~\label{fig:vis1}
\end{figure}

\subsubsection{Experimental setting}

The study was run in a closed workshop with 10 participants recruited from the University campus population. After a briefing on the TAPESTRY service, participants were invited to read the crowdfunding campaigns, browse the background description and biographies and invest a hypothetical \$1000 `TAPESTRY currency' between the eight campaigns. We randomly split the participant group into two; only one group was provided with TAPESTRY visualisations on the mock crowd-funding site (Fig.~\ref{fig:vis1}). Participants could use the mock site or wider resources on the Internet to help them make decisions to allocate the money. The study lasted 35 minutes; participants were asked to make one decision every 5 minutes given the knowledge they gleaned from their full use of Internet resources.  
\subsubsection{Experiment results}
\begin{figure*}[!ht]
\centering
    \includegraphics[width=\textwidth]{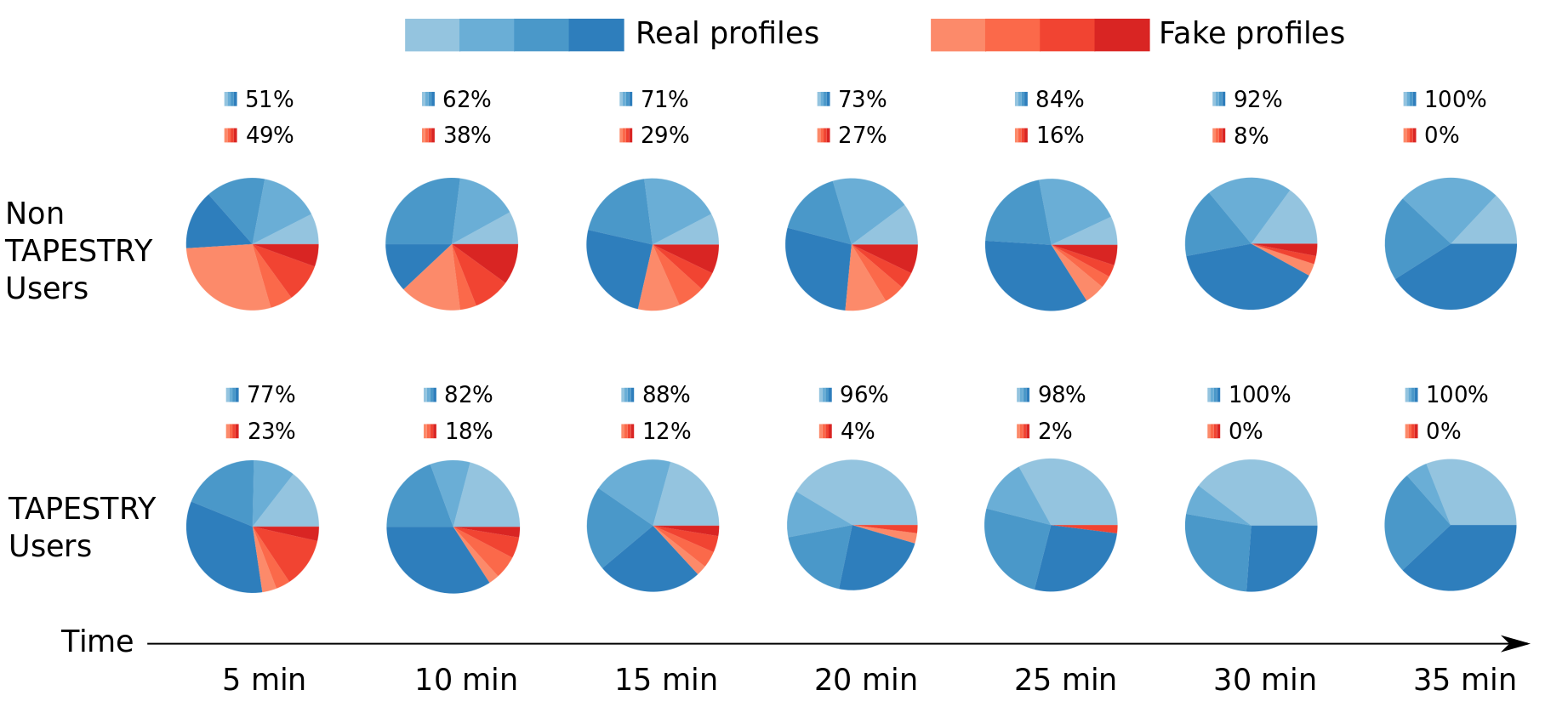}
        \caption{Participants' investment performance in the mock crowd-funding experiment:  top row is the control group without TAPESTRY visualisation; bottom row with TAPESTRY. The users with access to TAPESTRY achieved similar or better accuracy in detecting fake profiles, but did so at least twice as quickly.}~\label{fig:res}
\end{figure*}
We evaluated the participants' performance based on their investment results, comparing the amounts invested in real and fake profiles for both groups.  We consider investment in a fake profile (\ie scammer) a bad investment.  Fig.~\ref{fig:res} shows that  the accuracy of the investment results correlates to the time taken in background research; the more participants gathered information from their searches on the Internet, the more accurately they made their investment. Given the time limit, the TAPESTRY group used the visualisation tool to quickly understand the games developers' Twitter identity, speeding up their search to establish legitimacy.  We can conclude that although participants reached similar, correct decisions (in terms of discriminating their investment between genuine and fake developers) the time-to-task was considerable shorter (approx. by half) for TAPESTRY users.

\section{Conclusion}

We presented TAPESTRY; a novel Blockchain based service that enables users to determine the provenance of online identity from their digital personhood (DP) in order to make better decisions on who to trust online.  We  applied the TAPESTRY service to two tasks; determining the trustworthiness of another unknown individual (interpersonal trust), and determining the integrity of one's own social media feed (intraspective trust).  We used machine learning techniques to extract trust evidence from social media activities in a privacy preserving manner, and a proof-of-work Blockchain to store hashes of that evidence in order to underwrite its provenance.  Our service enabled users to then selectively disclose trust evidence to one another in order to prove the provenance of their identity.  To improve comprehension of the high volumes of evidence shared between users, we designed visualization techniques to summarise that evidence.  We evaluated the end-to-end system using a mocked up crowd-funding exercise run in a user workshop, and showed that TAPESTRY enables people to make accurate trust decision faster than the control group without access to the service.  We evaluated the end-to-end system for anomaly detection and showed that TAPESTRY enabled users to detect anomalies in a social media feed with accuracy of $\sim94\%$.  

Currently TAPESTRY is a prototype and future work will explore at-scale deployment beyond workshop settings. At scale it will become necessary to run multiple data lake services, with users distributed across different lakes.  This will add value to the PoW Blockchain which will then be maintained across multiple lakes.   At this stage further characterization of the performance of the TAPESTRY Blockchain should be undertaken.  Nevertheless we do not believe at-scale deployment of TAPESTRY is necessary to demonstrate the value in our hybrid on- and off-chain architecture for identity provenance, and the novel machine learning and visualization techniques developed for the service.

\ifCLASSOPTIONcompsoc
  \section*{Acknowledgments}
\else
  \section*{Acknowledgment}
\fi

TAPESTRY is funded by EPSRC Grant Ref: EP/N02799X/1 under the UKRI Digital Economy Programme.

\ifCLASSOPTIONcaptionsoff
  \newpage
\fi



%

\bibliographystyle{ieee}
\bibliography{main}

%

\begin{IEEEbiography}
[{\includegraphics[width=25mm,height=30mm,clip,keepaspectratio]{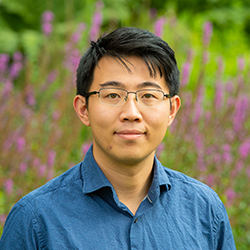}}]
{Yifan Yang}
 is a Research Fellow at the Centre for Vision Speech and Signal Processing (CVSSP), University of Surrey.  His research interests include but not limited to machine learning, image understanding and knowledge representation and reasoning.
\end{IEEEbiography}

\begin{IEEEbiography}
[{\includegraphics[width=25mm,height=30mm,clip,keepaspectratio]{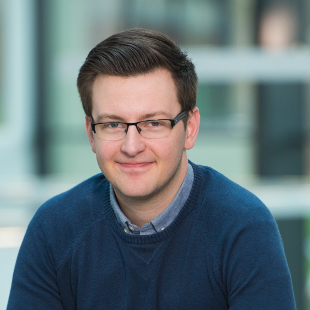}}]
{Daniel Cooper}
is a Research Software Developer at the Centre for Vision Speech and Signal Processing (CVSSP), University of Surrey. He manages the University of Surrey DLT testbed infrastructure, as well as the design and development of software, integrating AI technologies with DLT. 
\end{IEEEbiography}

\begin{IEEEbiography}
[{\includegraphics[width=25mm,height=30mm,clip,keepaspectratio]{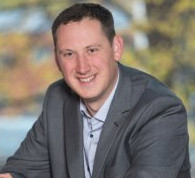}}]
{John Collomosse}
is a Professor at the Centre for Vision Speech and Signal Processing, University of Surrey.  His interests are in the fusion of Artificial Intelligence and Distributed Ledger Technology (DLT), and he directs the Surrey Blockchain testbed comprising several UKRI  funded projects in this area.
\end{IEEEbiography}
\begin{IEEEbiography}
[{\includegraphics[width=25mm,height=30mm,clip,keepaspectratio]{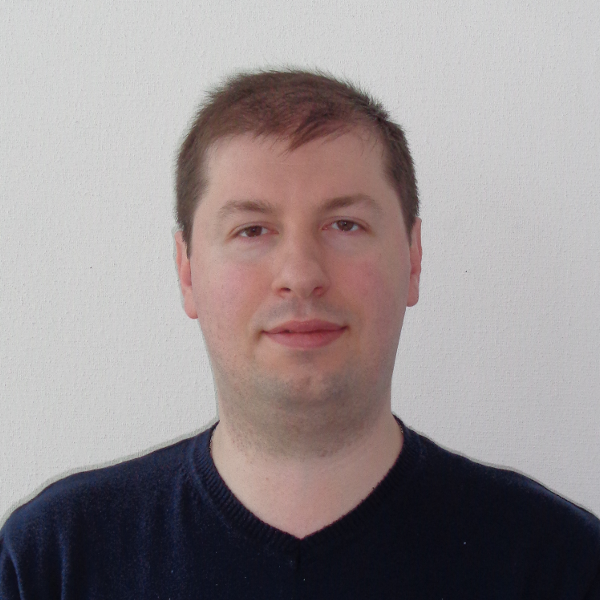}}]
{Constantin C\u{a}t\u{a}lin Dr\u{a}gan}
is a Research Fellow at the Surrey Centre for Cyber Security (SCCS) and Department of Computer Science, University of Surrey, UK. His research interests are in areas of applied cryptography, security analysis, formal methods, and privacy.  Previously, he worked as Research Fellow for CNRS \& INRIA, France.
\end{IEEEbiography}

\begin{IEEEbiography}
[{\includegraphics[width=25mm,height=30mm,clip,keepaspectratio]{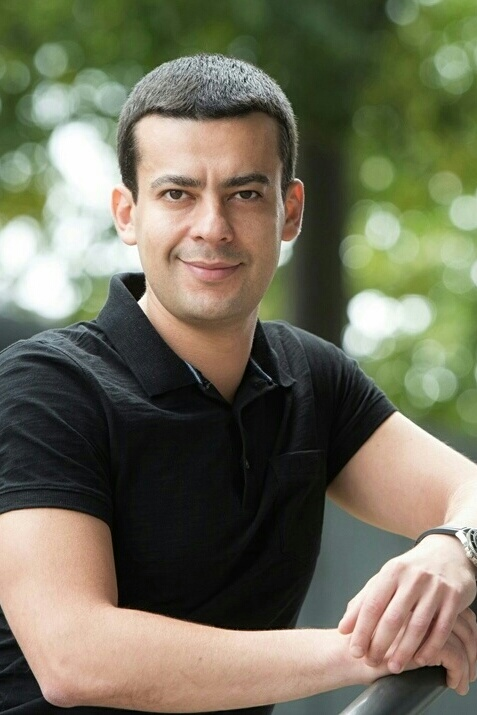}}]{Mark Manulis}
is Deputy Director of Surrey Centre for Cyber Security (SCCS) and Senior Lecturer in the Department of Computer Science, University of Surrey, UK. His research interests are in applied cryptography, network security and privacy. 
\end{IEEEbiography}

\begin{IEEEbiography}
[{\includegraphics[width=25mm,height=30mm,clip,keepaspectratio]{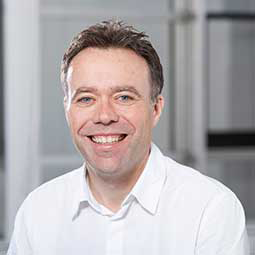}}]{Jamie Steane}is Associate Professor At Northumbria School of Design, Newcastle upon Tyne, UK. His design research lies at the intersection of design, business and digital technology and has involved the commercial development of early interactive products and services for the creative industries, financial and educational sectors.
\end{IEEEbiography}

\begin{IEEEbiography}
[{\includegraphics[width=25mm,height=30mm,clip,keepaspectratio]{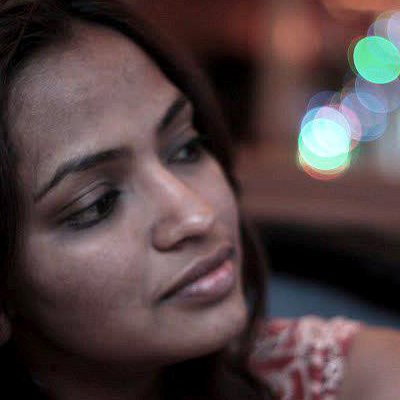}}]{Arthi Manohar} is a researcher and lecturer in the Department of Design, Brunel University London. Her research interests include participatory design, co-design, user-centered design and Human Computer Interaction. 
\end{IEEEbiography}

\begin{IEEEbiography}
[{\includegraphics[width=25mm,height=30mm,clip,keepaspectratio]{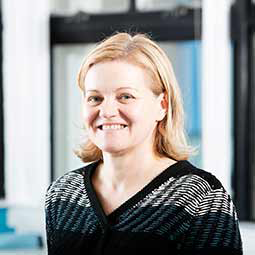}}]{Jo Briggs}
is Associate Professor at Northumbria School of Design, Newcastle upon Tyne, UK. Her research concerns designing tools for safer and enjoyable online interaction and investigations into and through the creative collaborative economy. She leads on design for a number of interdisciplinary projects and publishes on design and socio-technical subjects. 
\end{IEEEbiography}

\begin{IEEEbiography}
[{\includegraphics[width=25mm,height=30mm,clip,keepaspectratio]{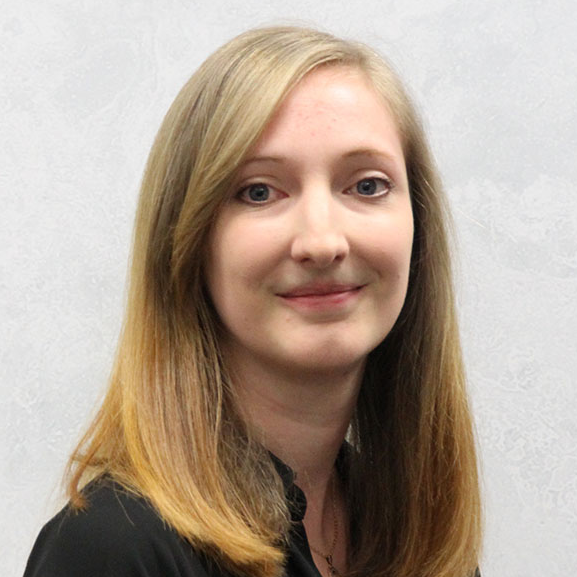}}]{Helen Jones} is a lecturer and researcher in Experimental Social Psychology at the University of Central Lancashire, UK. Her primary research interest lies in understanding how people behave online, in particular focusing on the decision-making processes that can leave them vulnerable to cyber security threats. 
\end{IEEEbiography}

\begin{IEEEbiography}
[{\includegraphics[width=25mm,height=30mm,clip,keepaspectratio]{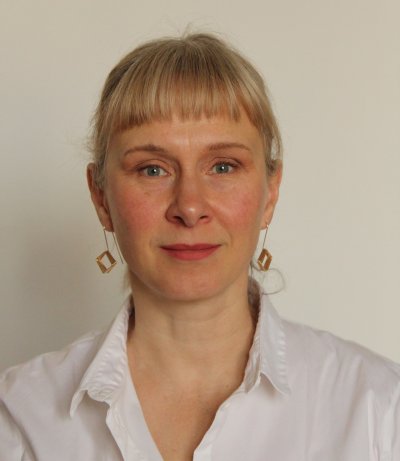}}]{Wendy Moncur}
is Interdisciplinary Professor of Digital Living at the University of Dundee. She leads the Living Digital group, which focuses on human factors in cybersecurity, online identity, personal data and delivering personal agency to users. 
\end{IEEEbiography}




\end{document}